\documentclass[12 pt]{article}

\usepackage{amsmath,amstext,amsgen,amsbsy,amsopn,amsfonts,graphicx,overcite,theorem}

\begin{document}

\title{A Graphical User Interface to Simulate Classical Billiard Systems}

\author{Steven Lansel \\ School of Electrical and Computer Engineering \\ School of Mathematics \\ gtg223g@mail.gatech.edu \\ \\ Mason A. Porter \\ School of Mathematics and \\ Center for Nonlinear Science, School of Physics \\ mason@math.gatech.edu}

\maketitle


\section*{Abstract}

Classical billiards constitute an important class of dynamical
systems.  They have not only been in used in mathematical 
disciplines such as ergodic theory, but their properties
demonstrate fundamental physical phenomena that can be observed
in laboratory settings.  This document provides instructions for a
Matlab module that simulates classical billiard systems.  It is
intended to be used as both a research and teaching tool. At present, 
the program efficiently simulates tables that are 
constructed entirely from line segments and elliptical arcs. It functions 
less reliably for tables with more complex boundary components.  The 
program and documentation can be downloaded from
\textit{http://www.math.gatech.edu/$\sim$mason/papers/}.

\section{Introduction}

In classical billiard systems, a point particle is confined to a
region in configuration space and collides with the boundary of
the region such that the angle of incidence equals the angle of
reflection.  As the velocity of the point particle is constant,
billiard systems are Hamiltonian.\cite{katok} Depending on the
geometry of a particular billiard table, there exist integrable
and/or chaotic regions in phase space.

\section{Overview of the Program}

The billiard simulation tool is a Matlab module with a Graphical
User Interface (GUI).  It is run by executing 'billiards'
in Matlab's command window while the files are in Matlab's path.
Users specify billiard tables by selecting from eight different
preprogrammed tables or creating their own. The initial position
and velocity (angle) of a trajectory are subsequently typed or specified by
clicking on a point in phase space.  The desired number of
iterations is also entered. After the program simulates the
resulting collisions, the data can be exported and analyzed. Each
time the point particle collides with the boundary, the position
and direction (i.e., momentum) are calculated. The position is
described by an arclength parametrization of the table, and the
direction is described by an angle measured with respect to the
horizontal angle.

The symbolic math toolbox, which contains the Maple kernel, is
required in order to run the billiard program.  This 
toolbox is used to take the derivative of the table boundaries
with the \textit{diff} command.  This GUI billiard simulator works with Matlab
releases 12 and 13.

\section{Bunimovich Mushroom}

\subsection{Entering the table}

In order to illustrate how to use the billiard simulator, we will
demonstrate an example step-by-step.  The table we use is shaped
like a mushroom\cite{mushroom}; it consists of a semicircular
region with a rectangular region extending from the base of the
semicircle.  Mushroom billiards are scientifically interesting,
as they constitute a generalization of the stadium billiard with 
a divided phase space in which some regions are chaotic and others
are integrable. The completely integrable semicircle and the
completely chaotic stadium billiards are mushrooms with limiting
values for the width of the stem.

One opens the billiard simulator by executing 'billiards' in the Matlab
command window when the folder containing the program's files are
in the path.  Because the mushroom is a preprogrammed table, it is
selected from the pop-up menu with 'Pick a table' as the default
selection.  A pull-down bar, four labels, and edit boxes will
appear. The pull-down bar is used to select either a circular or
elliptical mushroom.  In this example, we will work with a
circular mushroom.  Numbers are entered into the edit boxes to
select the desired dimensions of the mushroom. The 'radius' refers
to the radius of the semicircle, and the 'height' and 'width' of the
stem refer to the dimensions of the rectangular region of the
mushroom. The ratio of left to right sides enables the creation of
desymmetrized mushrooms and should be set to $1$ for symmetric
mushrooms.  After the four parameters have been entered, a preview
of the mushroom will be displayed.

\subsection{Entering the initial conditions}

\begin{figure}[htb]
\begin{centering}
\leavevmode
\includegraphics[width=3.6in, height=3in]{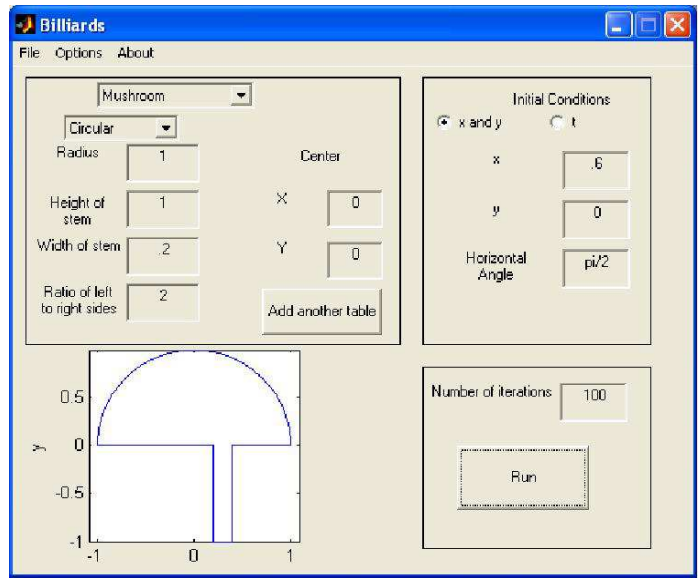}
\caption{Entering a mushroom table and initial conditions.}
\label{figgy1}
\end{centering}
\end{figure}

Once the table is created, one specifies the initial conditions
and the number of iterations.  Initial conditions can be entered
in two manners.  Click on the $x$ and $y$ button.  The $x$ and $y$
locations give the initial position of the point particle. The
angle specifies the initial direction of the point particle and is
measured in radians.  The 'number of iterations' specifies how
many collisions the simulator will calculate. The 'run' button
begins the simulation. Figure \ref{figgy1} shows a screen shot of the program
prior to running the simulation.

\subsection{Running the simulation}

When the 'run' button is pressed, Matlab begins the simulation.
The program displays the number of iterations completed out of the
total number requested.  The 'stop' button discontinues the
simulation. The module will not respond
 to user interactions until the current iteration is
completed.

\subsection{Analysis}

\begin{figure}[htb]
\begin{centering}
\leavevmode
\includegraphics[width=3.6in, height=3in]{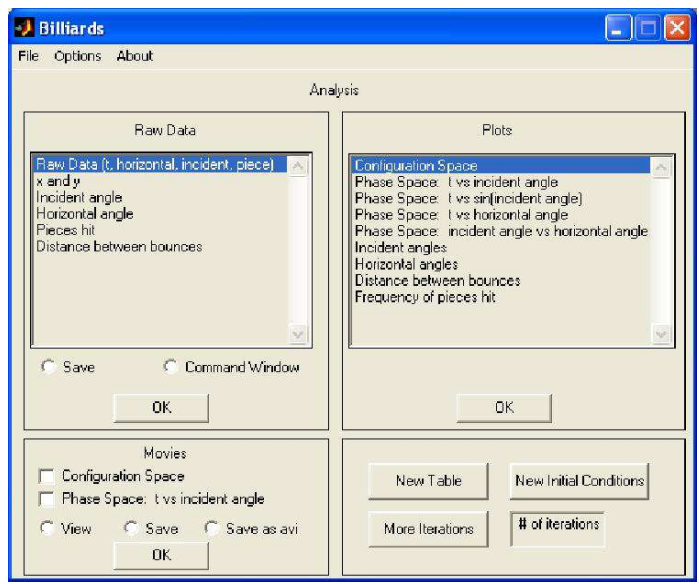}
\caption{Analysis window for Billiards.}
\label{figgy2}
\end{centering}
\end{figure}

The analysis window displayed in Figure \ref{figgy2} will appear after the
simulation is complete.  The raw data from the simulation can be
exported with the raw data box on the left. This option enables
the user to further analyze the calculated data. The desired type
of data should be selected from the list box on the left and can
either be displayed in the command window or saved by choosing the
appropriate radio button. The 'OK' button on the left exports the
data.  The 'Pieces hit' option gives the symbolic dynamics of the
initial conditions; the pieces are assigned numbers based on the
order of the parametric functions in the piecewise definition of
the table.

\begin{figure}[htb]
\begin{centering}
\leavevmode
\includegraphics[width=3in, height=3in]{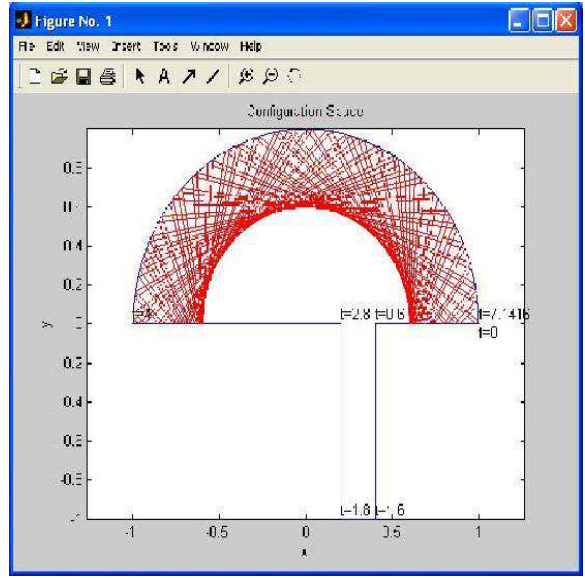}
\caption{Configuration space for the mushroom with an integrable
trajectory depicted.} \label{figgy3}
\end{centering}
\end{figure}

\begin{figure}[htb]
\begin{centering}
\leavevmode
\includegraphics[width=3in, height=3in]{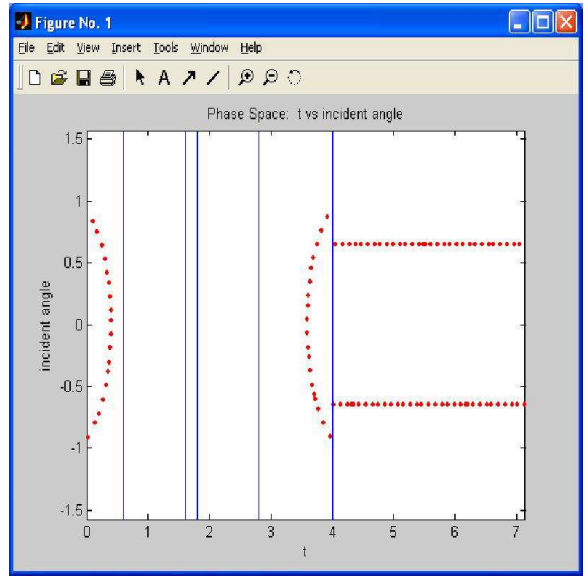}
\caption{Phase space for the mushroom with an integrable
trajectory depicted.} \label{figgy4}
\end{centering}
\end{figure}

Specific plots can be generated with the list box on the right. The
configuration space (see Figure \ref{figgy3}) displays the table and the
paths of any trajectories.  The phase space plots display Poincar\'e
sections of the billiard system.  The variable $t$ gives the
location on the boundary at which the point particle collides; it
is defined by the parametric equations describing the table.  The
incident angle $\phi$ gives the direction of the point particle
after the collision and is measured relative to the normal of the
boundary at the collision point.  The horizontal angle $\theta$
gives the direction of the point particle after the collision with
respect to the horizontal. Figure \ref{figgy4} shows the phase space of $t$
vs $\sin(\phi)$ for the Bunimovich mushroom. The incident angles, horizontal
angles, distance between bounces, and frequency of pieces hit
options display histograms of the appropriate data, although it
may be desirable to export this data in order to perform
additional analysis. The histograms are generated using the Matlab
function \textit{hist}. The plots are displayed by pressing the
'OK' button on the right side. The billiard table and data can be
saved and opened later using the file menu.

\subsection{New initial conditions}

The options in the bottom right of the window enable the user to
generate additional data.  The 'New Table' button clears
all data and resets the program so that a new table can be
entered. The 'New Initial Conditions' button retains all the
calculated data, and the user can enter new initial conditions for
the current table.  More iterations are calculated for the same
initial conditions by entering the desired number of additional
iterations and pressing the 'More Iterations' button. The program
will continue where the last simulation stopped.

\begin{figure}[htb]
\begin{centering}
\leavevmode
\includegraphics[width=3.6in, height=3in]{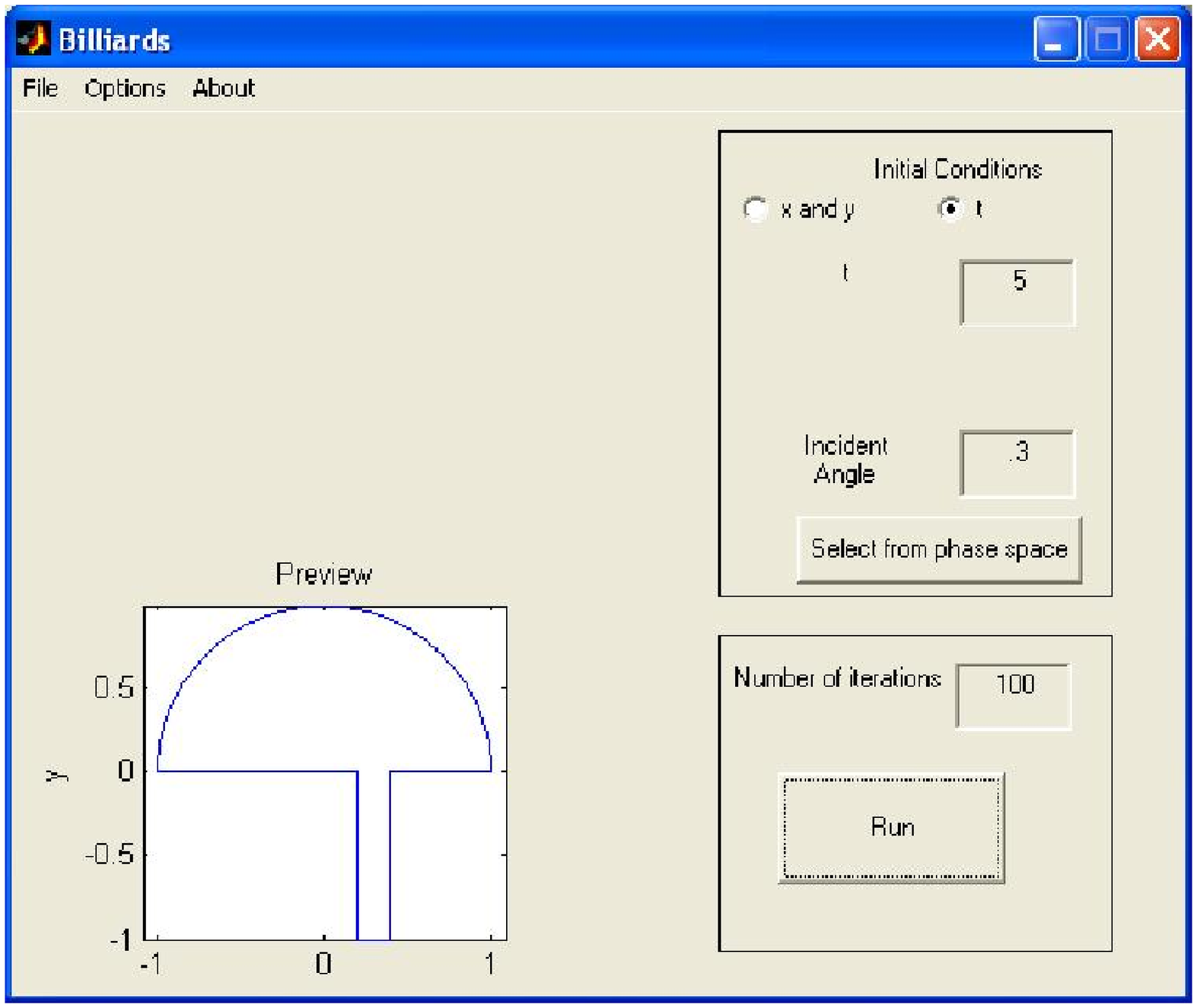}
\caption{Entering new initial conditions for the mushroom.}
\label{figgy5}
\end{centering}
\end{figure}

\begin{figure}[htb]
\begin{centering}
\leavevmode
\includegraphics[width=6in, height=3in]{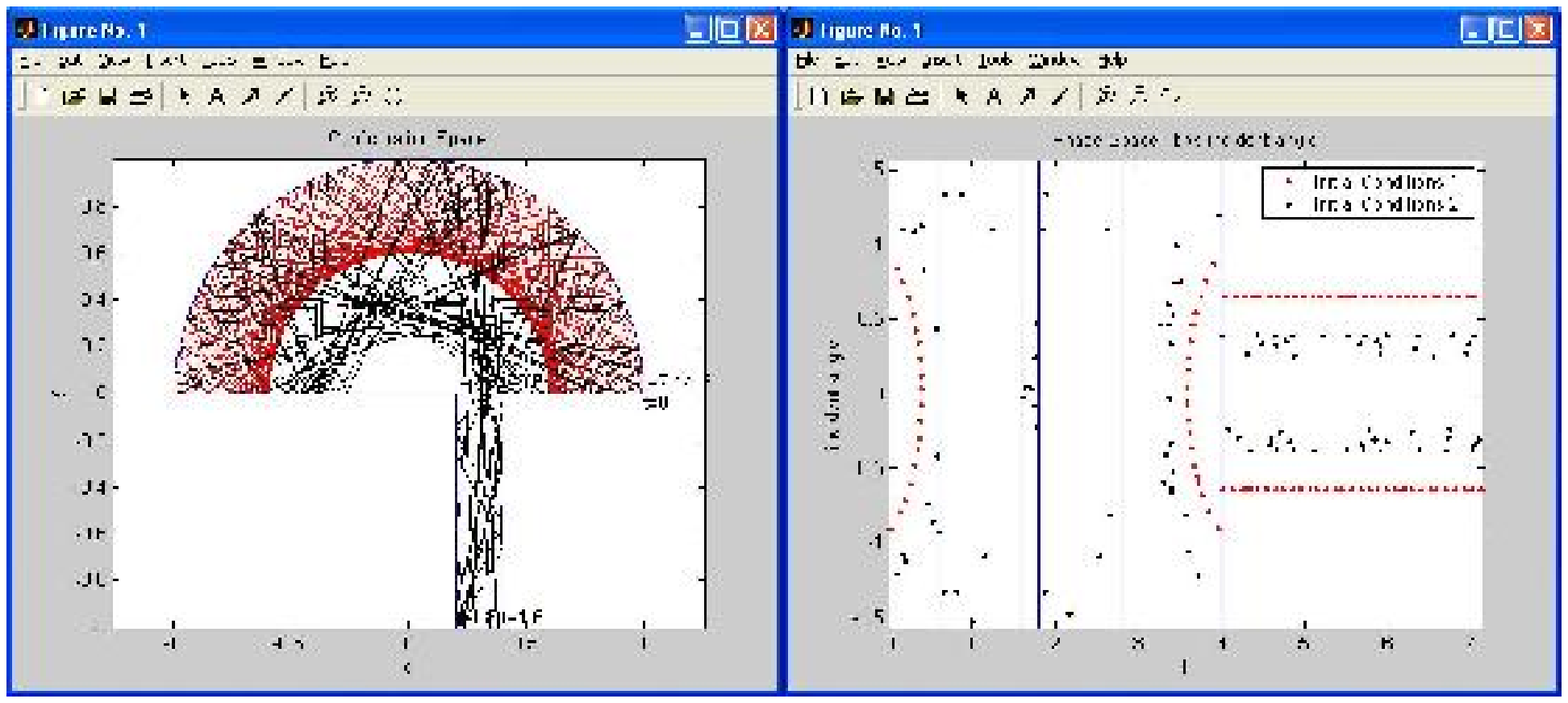}
\caption{Configuration space and phase space of a mushroom billiard with two 
initial conditions depicted.} \label{figgy6}
\end{centering}
\end{figure}

To continue the mushroom example, click on the 'New Initial
Conditions' button.  This time, we will enter the initial
conditions using the arclength variable $t$.  With this method, one 
specifies a location
on the table boundary with $t$ and a direction by specifying the
incident angle of the trajectory to the table.  One may either
type in values for $t$ and the incident angle or select them from
phase space.  Clicking on the 'Select from phase space' button
will open a plot of phase space with data from all previous
initial conditions.  Use the cross hairs to select an initial
condition, and the corresponding coordinates are entered into the
billiard simulator.  If more accuracy is desired than is possible
with the cross hairs, the zoom tool can be used to magnify a
particular portion of phase space.  In this case, the coordinates from phase
space must be typed into the program. To continue with this
guided example, enter the initial conditions shown in Figure \ref{figgy5}.

After Matlab's computations are done, the analysis window is again
displayed. The resulting configuration and phase space plots are
depicted in Figure \ref{figgy6}. Note that the first set of initial
conditions specifies a trajectory that remains in the semicircular
region of the mushroom.  The set of all such trajectories comprise
an integrable region of phase space. On the other hand, the second
set of initial conditions corresponds to a trajectory that
collides with the stem of the mushroom. These trajectories form a
chaotic region of phase space. This example shows how circular
mushrooms exhibit a divided phase space that contains exactly one
integrable region and exactly one chaotic region.\cite{mushroom}

\section{Composite Billiard Tables}

\begin{figure}[htb]
\begin{centering}
\leavevmode
\includegraphics[width=3.6in, height=3in]{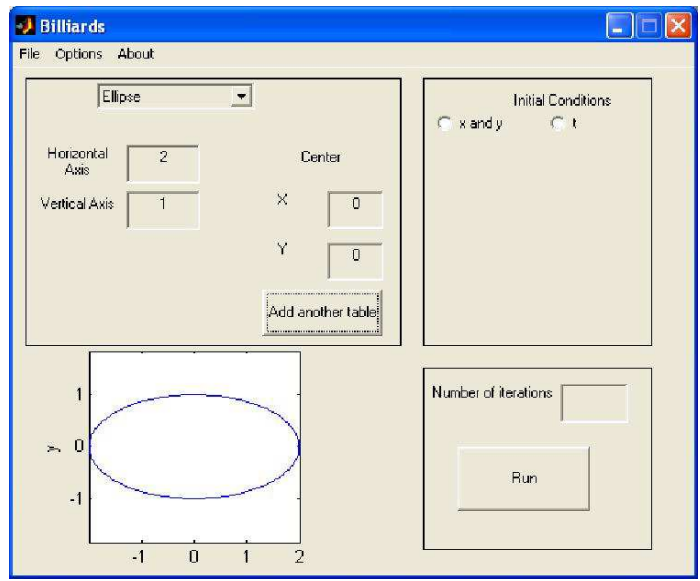}
\caption{Entered parameters and preview of the ellipse.}
\label{figgy7}
\end{centering}
\end{figure}

Composite billiard tables can be constructed by combining multiple
simple billiard tables.  Such composite tables are necessary, for
example, when considering any table in which the boundary cannot be
described by one continuous curve.  (One composite table, the Sinai billiard, 
is preprogrammed.)  One can implement such tables with the 'Add another table' 
button.  To illustrate this, consider a composite table consisting of an
off-center circle inside an ellipse.  First, the ellipse is
selected from the 'Pick a table' pop-up menu. One enters
parameters for the lengths of the horizontal and vertical axes.
The 'Add another table' button appears once both parameters are
entered. Figure \ref{figgy7} displays the entered ellipse and the 'Add
another table' button.

\begin{figure}[htb]
\begin{centering}
\leavevmode
\includegraphics[width=3.6in, height=3in]{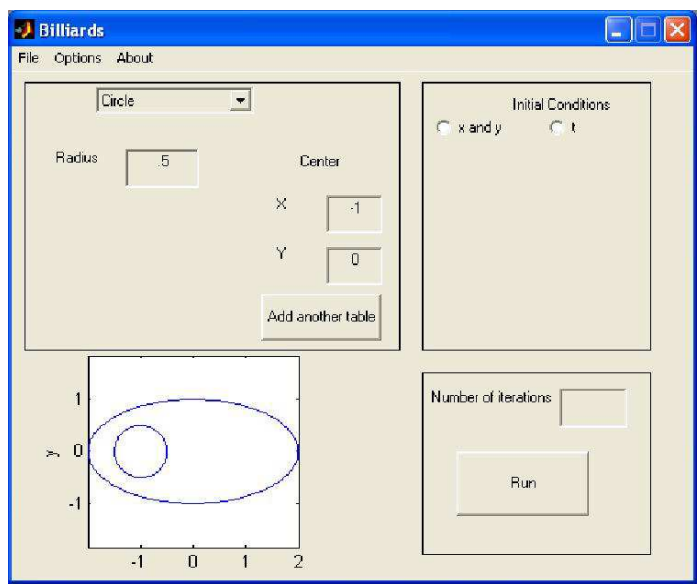}
\caption{Completed composite table.} \label{figgy8}
\end{centering}
\end{figure}

The second component of the composite table can now be added. It is entered as 
normal except that the previous
component remains a part of the table. In this example, the circle
option is selected from the pop-up menu. The radius is entered as
normal.  Now we need to move the circle so that the circle and the
ellipse are not concentric. This is accomplished by editing the
$x$ and $y$ coordinates under the center label.  The preview is
redrawn once the center of the current component of the table has
been moved to the new coordinates.  Figure \ref{figgy8} shows the completed
composite table.  The numbering of the pieces and value of the
variable $t$ for each subsequent component of a
composite table continue where the previous components left off.

\section{Billiard Table Maker}

The strength of the billiard program lies in its
generality, as it can simulate any classical planar billiard system. If
one desires to analyze a table that is not preprogrammed, it can
be designed using the Billiard Table Maker, which is opened by
selecting 'Custom table...' from the 'Pick a table' pull-down bar.

\begin{figure}[thb]
\begin{centering}
\leavevmode
\includegraphics[width=3.75in, height=3in]{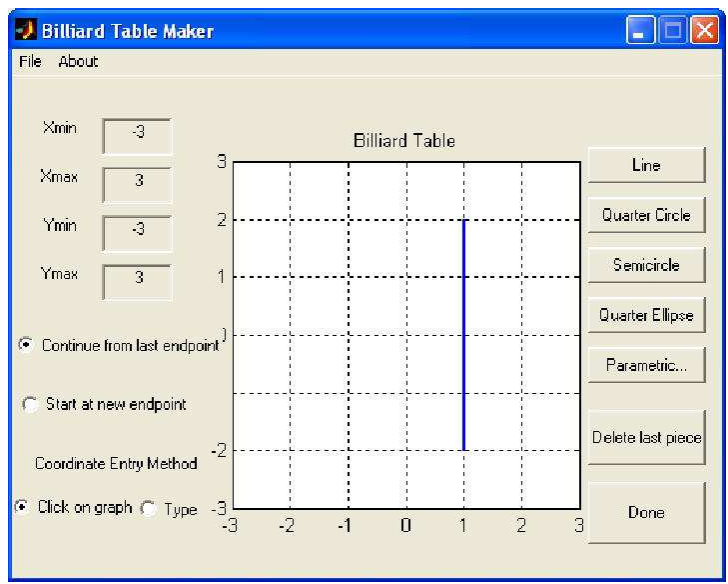}
\caption{Line segment drawn with the Billiard Table Maker.}
\label{figgy9}
\end{centering}
\end{figure}

As an example, we will use the Billiard Table Maker to create a
stadium billiard modified so that one of the straight segments is
replaced by a sinusoidal function.  We start by creating a
vertical line that will form one side of the modified stadium.  
To do this, click the 'Line' button.  At this point, note that the mouse
controls a pair of cross hairs that are used to select the
starting and ending points of the line.  For this example, $(1,2)$ and then
$(1,-2)$ were selected.  The program then draws the line segment
connecting these two points, as shown in Figure \ref{figgy9}.

\begin{figure}[htb]
\begin{centering}
\leavevmode
\includegraphics[width=3.75in, height=3in]{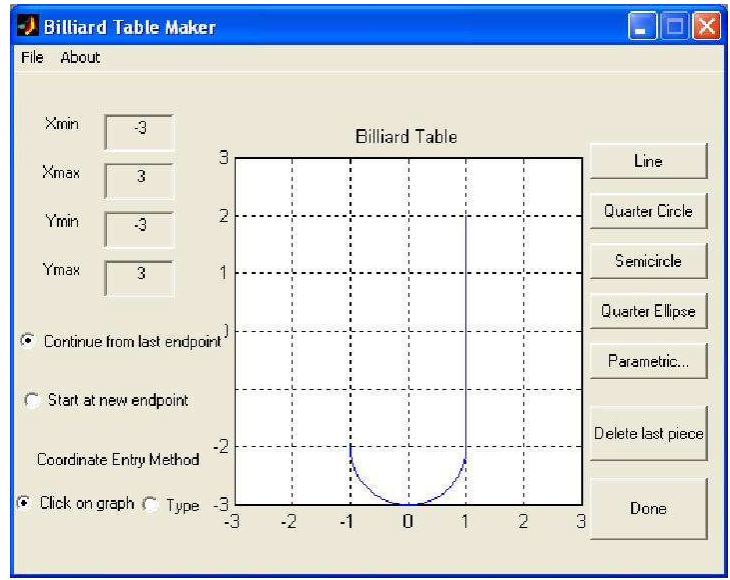}
\caption{Semicircle added to the endpoint of the previous piece.}
\label{figgy10}
\end{centering}
\end{figure}

We now wish to add a semicircular region to form the bottom of the
modified stadium.  Because we are adding a new piece where the
previous one ended, we want to use the 'Continue from last
endpoint' feature on the left, which is the default setting. The
'Start at new endpoint' feature allows one to construct tables that
cannot be described with a single boundary curve.  Click on the
'Semicircle' button.  Using the cross hairs, select $(-1,-2)$ for
the endpoint of the semicircle and select \textit{any} point above
the line connecting the endpoints of the semicircle to designate
the inside of the semicircle.  The resultant table is shown in
Figure \ref{figgy10}.

\begin{figure}[htb]
\begin{centering}
\leavevmode
\includegraphics[width=2.4in, height=1.5in]{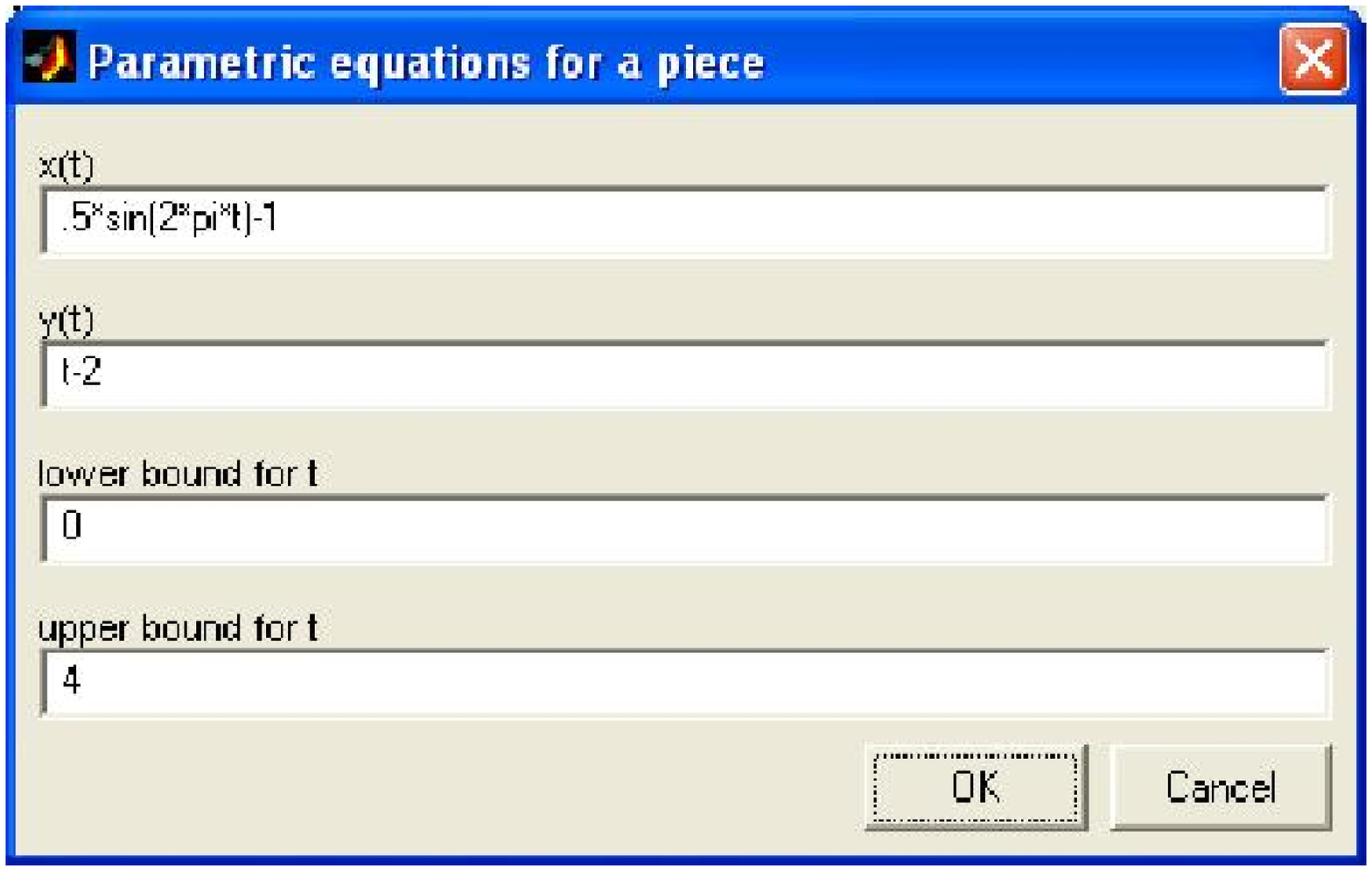}
\includegraphics[width=2.8in, height=2.25in]{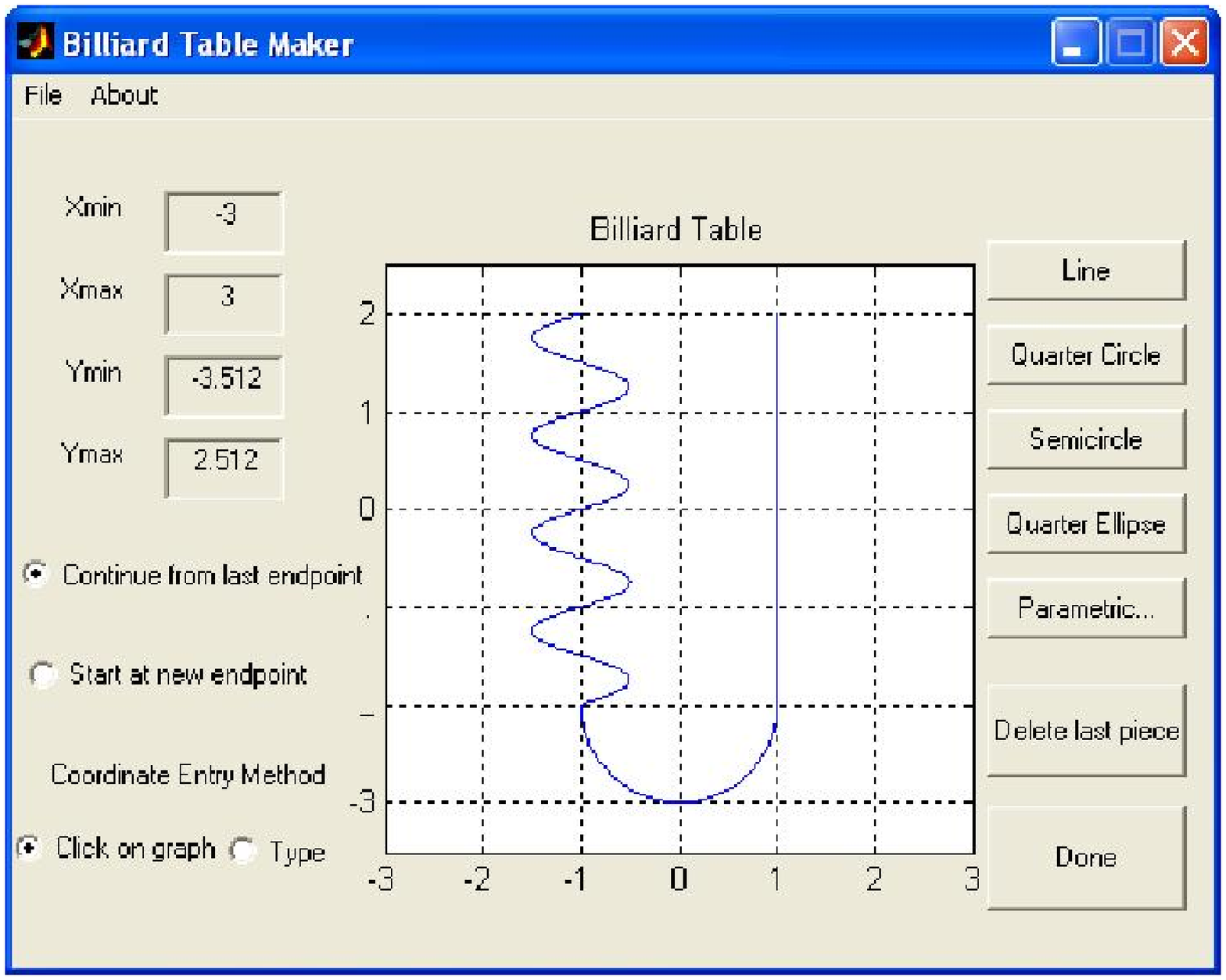}
\caption{Parametric equations for the sinusoidal piece of the
table and the resultant table.} \label{figgy11}
\end{centering}
\end{figure}

The next piece to add is a sinusoid for the left side of the
modified stadium.  As this is not a basic piece (line
segment, quarter circle, semicircle, or quarter ellipse), we must
use the 'Parametric...' push button.  With this feature, one can
add a piece by entering parametric equations of a curve. Clicking
on the push button brings up a window in which one enters $x(t)$,
$y(t)$, and lower and upper bounds for $t$.  Note that the default
value for the lower bound is $0$, but this can be changed to any
desired value. The parametric equations for this example and the
resultant table are shown in Figure \ref{figgy11}.  The vertical
axis of the window has automatically been scaled to make sure the
last piece is shown in the grid.  The axis can be edited manually by
changing the minimum and maximum values for $x$ and $y$ displayed
on the left.

\begin{figure}[htb]
\begin{centering}
\leavevmode
\includegraphics[width=2.4in, height=1.5in]{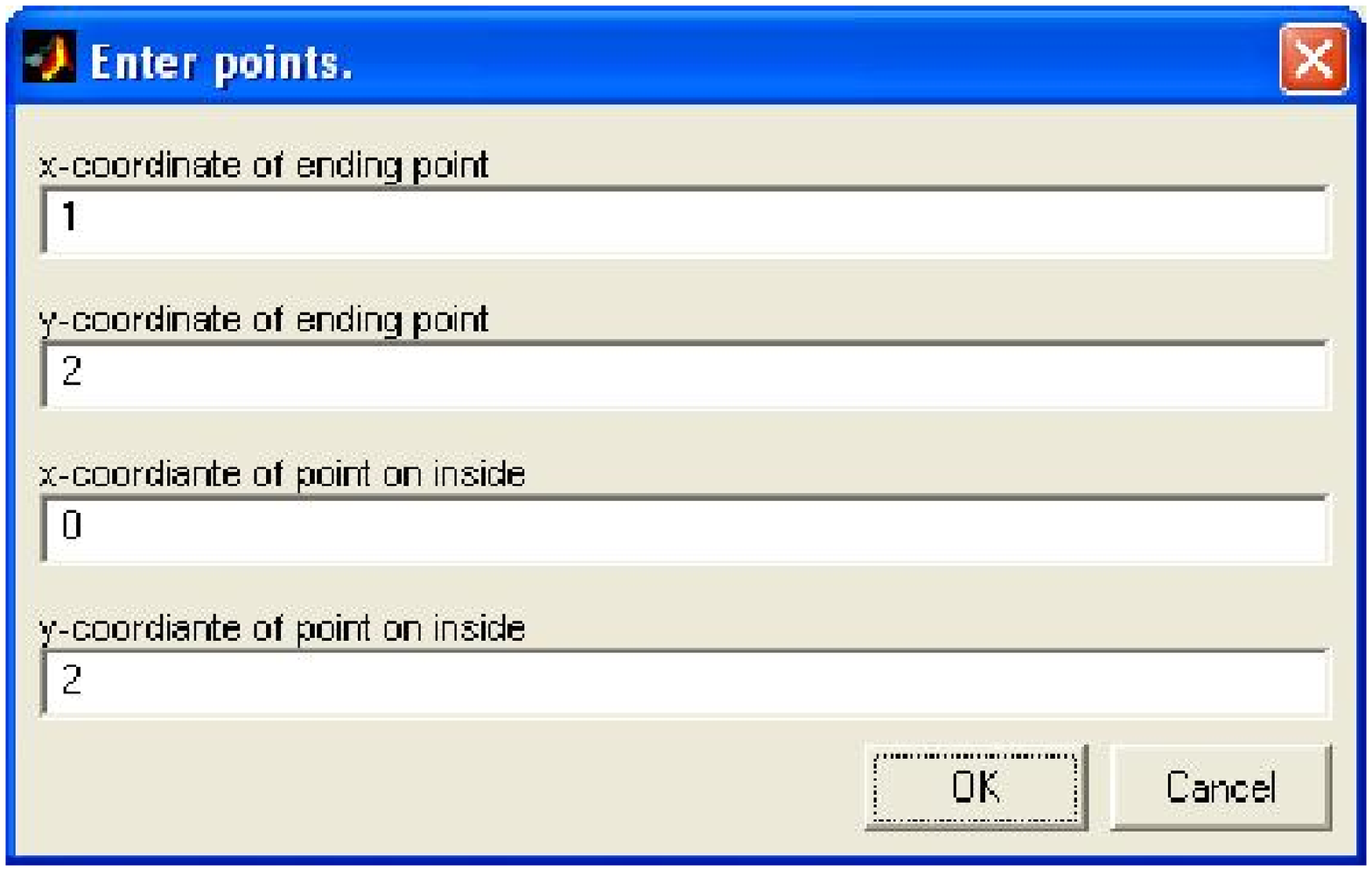}
\includegraphics[width=2.8in, height=2.25in]{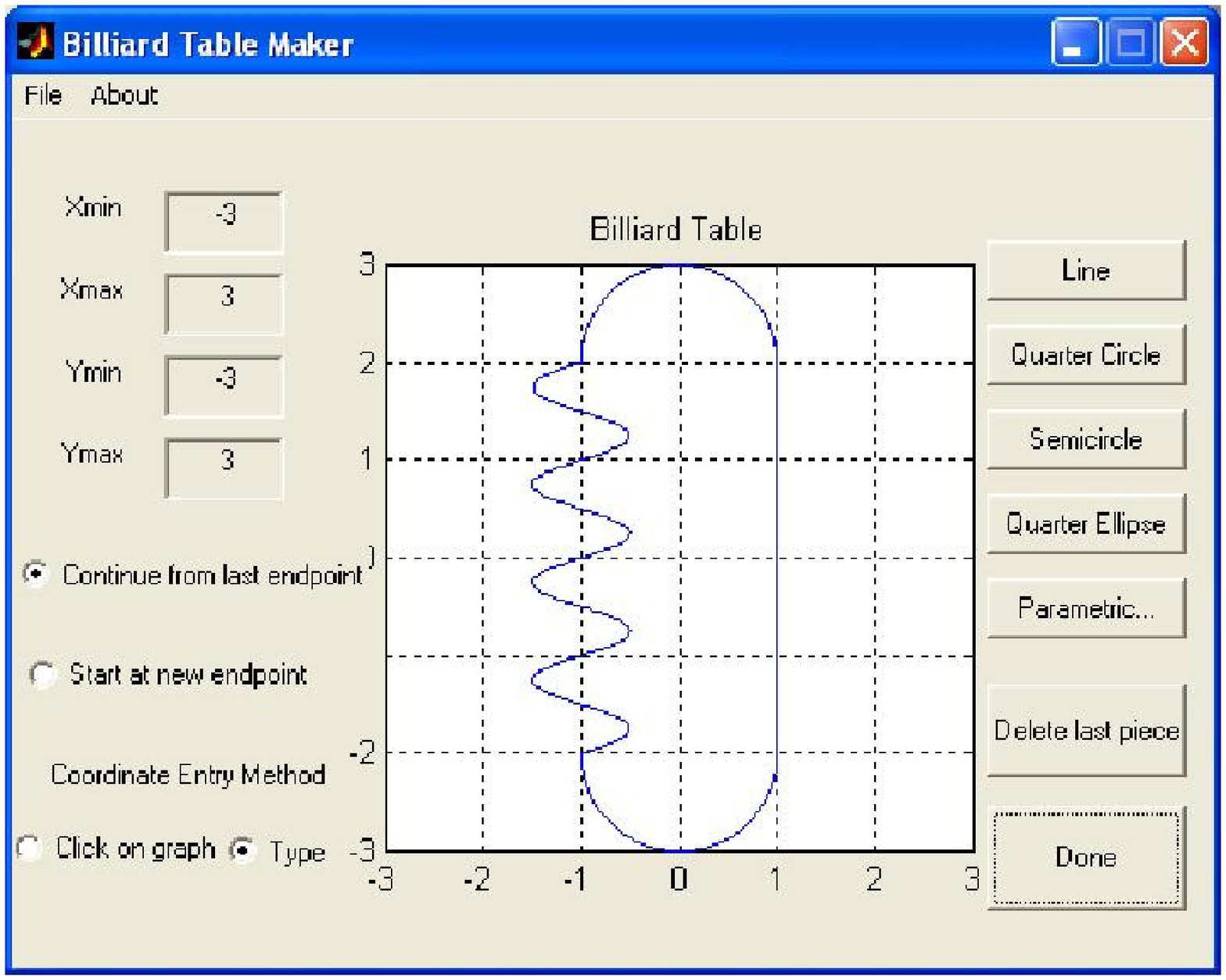}
\caption{Entering coordinate points for semicircle and completed
billiard table.} \label{figgy12}
\end{centering}
\end{figure}

We will now add an upper semicircle to complete the billiard
table. To illustrate this feature, select the radio button 'Type'
under 'Coordinate Entry Method.'  This enables the user to
accurately specify the coordinates of the input points. Because
the program rounds input points, the alternate method of clicking
on the graph to select points only allows points with integer
coordinates to be selected. Clicking on the 'Semicircle' push
button brings up a window to enter coordinates. The entered
coordinates and the final billiard table are shown in Figure \ref{figgy12}.
Quarter ellipses in which the major and minor axes are parallel to
the coordinate axes can also be created in a similar manner by
using the 'Quarter Ellipse' button. One can either save the table
now or return to Billiards by clicking on the 'Done' button.

\section{Movies}

One can create movies to view an animation of generated billiard
data.  Controls for doing this are located at the bottom left of the
analysis window.  Movies of configuration and/or phase space can
be created by checking the appropriate boxes.  Such movies show one
frame for each iteration and highlight the most recently drawn
iteration.  The frames rate can be modified using the options
menu.  Movies can only be created for the most recent initial
condition and should be used only with a small number of
iterations due to memory limitations. By
selecting the appropriate radio button, the movie can be viewed in
Matlab, saved as a Matlab movie, or saved as an .avi file.  For
every option, each frame of the animation is first displayed on
the screen. After all frames have been rendered, the movie will be
displayed or a save dialog box will appear depending on which
radio button is selected.

\section{Description of Tables}

The program stores billiard tables in a cell array.  For example,
Table \ref{tab1} shows the table created in the previous section.  A table
is given by a piecewise function; each row describes a piece
of the table. The first and second columns are parametric
equations for the $x$ and $y$ coordinates for each piece of the
table [$x(t)$ and $y(t)$]. The equations are stored as Matlab
inline functions.  The third and fourth columns give the bounds
for $t$, which is used as the dummy variable for the
parametrization. The functions $x(t)$ and $y(t)$ have been scaled
to the appropriate expressions so that $t$ also represents the arc
length from the starting point of the billiard table to the
current point for a given connected component of the billiard. For
composite tables, the value of $t$ for each component starts at
the same value of $t$ used at the end of the previous component.
The final column of the table contains a flag that determines the
nature of the piece. For line segments, $1$ is stored
in the final column; for circular or elliptical arcs, $2$
is stored in the final column; if the piece is neither a line segment
nor one of arcs specified above, then $0$ is stored in the final column.  
Note, however, that if a piece is entered by typing in parametric equations, 
then the final column is automatically $0$ regardless of the actual identity 
of the piece.  Circular and elliptical arcs that sweep out arbitrary angles 
must be entered parametrically.  The three different types of pieces are 
treated differently during the simulation.

Parametric equations that constitute exterior boundaries to the table
must be traced in the clockwise direction, whereas interior
boundaries must be traced in the counterclockwise sense.  For
example, the exterior square of the Sinai billiard is oriented clockwise,
and the interior circle is oriented counterclockwise.  If the
wrong direction is used to parameterize the boundary, specifying
initial conditions using $t$ and the incident angle will not work
properly, as the incident angle will be on the wrong side of
the curve.

\begin{table}
\centerline{
\begin{tabular}{|c|c|c|c|r|} \hline
$x(t)$ & $y(t)$ & lower bound & upper bound & flag \\ \hline
 1 & $2-t$ & 0 & 4 & 1 \\
 cos(-$t$+4) & -2+sin(-$t$+4) & 4 & 7.1416 & 2 \\
 .5*sin(2*$\pi$*($t$-7.1416))-1 & $t$-9.1416 & 7.1416 & 11.1416 & 0 \\
 cos[(-$t$+11.1416)+3.1416] &  2+sin[(-$t$+11.1416)+3.1416] & 11.1416 & 14.2832 & 2 \\
\hline
\end{tabular}}
\caption{Representation of table created in Section 5.}
\label{tab1}
\end{table}

\section{Description of Raw Data}

\begin{figure}[htb]
\begin{centering}
\leavevmode
\includegraphics[width=3in, height=3in]{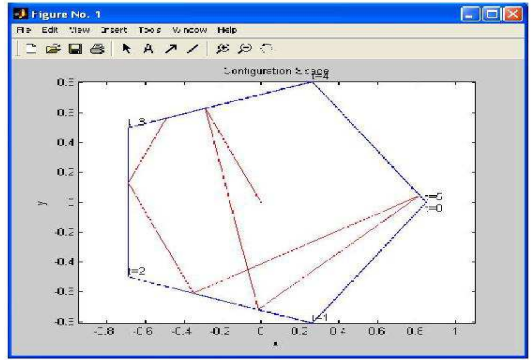}
\caption{Configuration space of pentagon simulation.}
\label{figgy13}
\end{centering}
\end{figure}

The billiard simulator calculates and stores information describing each
collision of the point particle with the boundary.  For example,
consider the simulation of a regular pentagon with side length
$1$ and initial location $(0,0)$ with an angle of $2$.  Figure \ref{figgy13}
shows the configuration space for this simulation; raw data is
presented in Table \ref{tab2}.

\begin{table}
\centerline{
\begin{tabular}{|c|c|c|r|} \hline
$t$ & horizontal angle & incident angle & piece \\ \hline
 3.4205 & -1.3717 & -0.1150 & 4.0000 \\
 1.2935 &  0.7434 & -0.5133 & 2.0000 \\
 4.9418 & -2.6283 & -0.1150 & 5.0000 \\
 1.6438 &  2.0000 &  0.7434 & 2.0000 \\
 2.6301 &  1.1416 &  1.1416 & 3.0000 \\
 3.2091 & -0.5133 &  0.7434 & 4.0000 \\
\hline
\end{tabular}}
\caption{Data for pentagon simulation.} \label{tab2}
\end{table}

Each collision with the table is described by a row of data.  The
location of the point on the boundary that the particle hits is
given in the first column as a value of $t$. This value can be
converted into $x$ and $y$ coordinates by evaluating the
appropriate expressions in the table matrix.  The second column
gives the horizontal angle $\theta \in [-\pi,\pi]$, which is a
measure of the angle of a vector in the direction the particle
travels after colliding with the boundary. The quantities $t$ and
$\theta$ give, respectively, the position and direction of the
point particle immediately after its collision with the boundary.
The third column contains the tangential angle
$\phi \in [-\pi/2,\pi/2]$, which is the angle between the normal to
the position of the boundary the particle hits and the exiting
path taken by the particle. Negative angles indicate the exiting
path is clockwise from the normal, whereas positive angles
indicate that the exiting path is counterclockwise from the
normal. The fourth column gives the piece the particle hits. The
sequence of such pieces encodes a symbolic dynamics for the given
trajectory. Sets of data for each initial condition are stored by
the program in a cell array called 'data.'

\section{Calculation of iterations}

The program runs iteratively in its simulation of classical
billiards.  Given the position and direction of the previous
collision, the program calculates the position and direction of
the point particle after its subsequent collision with the
boundary.  To find the location of the next collision, the
program searches for an intersection between the line that
describes the path of the point particle and each of the table's
parametric pieces.  Given all such intersections, the point with
the minimum distance traveled is the next point of intersection.
To find the direction in which the point particle travels
following the collision, the normal angle to the boundary is
computed from the derivatives of the equations of the table at the
point of intersection.  Addition and subtraction of angles is used
to calculate the exit angle from the normal angle and the entrance
angle:
\begin{align}
    \theta_{n}&=2\arctan\Big{(}\frac{dy}{dt}\Big{/}\frac{dx}{dt}\Big{)}\Big{|}_{t_{n}}-\theta_{n-1}\,,
    \notag
    \\
    \phi_{n}&=\arctan\Big{(}\frac{dy}{dt}\Big{/}\frac{dx}{dt}\Big{)}\Big{|}_{t_{n}}-\theta_{n-1}+\pi/2\,, \label{atan}
\end{align}
where $\theta_{n}$ represents the angle with respect to
the horizontal of the $n$th iteration, $\phi_n$ represents
the incident angle of the $n$th iteration, $y(t)$ and $x(t)$ are the
parametric equations of the boundary, and $t_{n}$ is the value of
$t$ that gives the location of the $n$th intersection with the
boundary.  In Matlab, equation (\ref{atan}) is implemented using the 
`arctan2' function to ensure that one obtains the correct quadrant for the 
angle.  Additionally, note that $\phi_{n-1}$ does not appear in the 
right-hand-side of (\ref{atan}), as this angle is used only for phase 
space plots and is not involved in the calculation of angles in subsequent 
iterations.


\subsection{Corners}

Special consideration must be employed if the point particle
collides with a corner of the billiard table, as such points correspond to 
singular points of the billiard (Poincar\'e) map obtained from examining only 
the collisions (and not the straight paths between them) of the vector field 
describing the billiard system.\cite{katok}  This occurs when
the point particle's path reaches a point where two pieces of
the table join abruptly (with discontinuous first derivative with
respect to arc length).  In the present numerical implementation, whenever 
the point particle collides with the boundary at a point where $t$ is 
within $10^{-8}$ of the beginning or end of a piece, it is considered to have
hit a corner. In order to numerically compute the angle with which the point
particle leaves the collision, the tangential angles of the two
pieces of the table are averaged. The point particle subsequently bounces
off a boundary oriented at this angle as if the collision were
normal.  (When studying billiard systems using this program, one needs to be 
careful if a trajectory hits the boundary too close to a corner.)

\section{Precision}

Errors due to round-off can grow quickly with our billiard
simulations, as frequently occurs for repetitive numerical
approximations. The rate that the errors compound depends
fundamentally on the geometry of the billiard table.  For certain
billiard tables, the error is negligible for a very large number
of iterations. For others, this is not the case.

Two examples are presented to demonstrate how the accuracy of the
simulation depends on the particular table.  In the circular billiard 
with unit radius, we examined the trajectory starting at $(.5,0)$ with 
an initial horizontal direction.  The maximum error in the incident angle
after $10,000$ iterations was only $1.1213\times10^{-13}$.

Consider, however, a billiard table consisting of two circles of
unit radius with respective centers at $(1.5,0)$ and $(-1.5,0)$. The
initial conditions were set to the origin with a horizontal angle.
 The point particle was then calculated to escape the two circles due to
round-off error after $10$ iterations.  This extreme example of a numerically
unstable periodic orbit demonstrates that one must be cautious when using
the program.

\section{Known Errors}

The current program is not yet reliable for tables that contain
pieces that are not line segments or elliptical arcs.  The program
will work properly until it fails to find where a trajectory
intersects a complex curve.  This problem is due to an inability of
the program to reliably find a zero of a given function on a given
interval.

\section{Additional Features to be Implemented}

A useful analytical tool would entail the creation of a Markov
partition of phase space.\cite{predrag}  For each point in phase
space, the piece against which the point particle will collide on the
next iteration will be determined.  This will create a Markov
partition by dividing phase space so that every point in the same
region will collide with the same piece on the next iteration.
This will be visually implemented by coloring phase space. The
partition will allow one to easily view the symbolic dynamics for
the first several iterations associated with any initial condition.

One can implement a Markov partition by finding the critical
angles that cause the next collision of the point particle to
collide with a different piece. The vertical segment of phase
space that corresponds to this location can then be colored based
on which piece will next be hit. This process will then continue
for the entire boundary of the billiard table; all of
these vertical segments are then joined to form the Markov
partition.

If the regions of phase space described above can be found, the
Markov partition can be used to speed up the billiard simulator 
considerably.  For each iteration, the program currently must check 
all pieces of the table to find the proper intersection.  This process 
is very costly, as a floating point root finder is used each time to 
find all possible intersections between the point particle's path and 
the table.  With the implementation of a Markov partition, the program 
will be able to determine which segment the particle
subsequently hits instead of searching for the right piece. The progrom 
will no longer spend time finding irrelevant roots, which will improve 
the speed and efficiency of the billiard simulator substantially.

Future  versions of the program can also include computations of
important quantities such as Lyapunov exponents.  In the long run,
we would ultimately like to expand the program to simulate
quantum billiards as well as classical ones.  Phenomena such as scarring 
would then be especially easy to study.  


\section{Conclusions}

We created a GUI Matlab module that simulates classical billiards. It
provides a useful research and teaching tool for scientists interested in
 these dynamical systems.  This module's simulations can be used not only to 
produce accurate phase space and Poincar\'e section plots for scientific 
publications but also to gain considerable mathematical and physical insight.

\section*{Acknowledgements}

We gratefully acknowledge Leonid Bunimovich for useful scientific discussions 
during this research project and Peter Mucha for providing assistance with 
technical Matlab issues.  The REU summer program at the School of Mathematics
 and the President's Undergraduate Research Award (PURA) at the Georgia
Institute of Technology provided financial assistance.


\end{document}